\documentclass{icrc2009}

\usepackage{graphicx}   
\usepackage[caption=false]{caption}    
\usepackage[font=footnotesize]{subfig} 
\usepackage{fixltx2e}
\usepackage{url}

\newcommand{\shorttitle}[1]%
{\markboth{Proceedings of the 31\MakeLowercase{$^{st}$} ICRC, {\L}\'{o}d\'{z} 2009}{#1} }
\newcommand{\etal}{\MakeLowercase{\textit{et al. }}} 

\begin{document}
\title{A new surface parameter for composition studies at high energies}

\author{
\IEEEauthorblockN{G. Ros\IEEEauthorrefmark{1},
		 A. D. Supanitsky\IEEEauthorrefmark{2},
                 G. A. Medina-Tanco\IEEEauthorrefmark{2},
                 L. del Peral\IEEEauthorrefmark{1} and
                 M. D. Rodr\'iguez Fr\'ias\IEEEauthorrefmark{1}}

\IEEEauthorblockA{\IEEEauthorrefmark{1}Space Plasmas and AStroparticle Group, Dpto. F\'isica, Universidad de Alcal\'a. Ctra. Madrid-Barcelona km. 33. \\ Alcal\'a de Henares, E-28871 (Spain).}
\IEEEauthorblockA{\IEEEauthorrefmark{2}Instituto de Ciencias Nucleares, UNAM, Circuito Exteriror S/N, Ciudad Universitaria, \\ M\'exico D. F. 04510. (Mexico).}
}

\shorttitle{G. Ros \etal New parameter for composition}
\maketitle

\begin{abstract}

A new family of parameters intended for composition studies is presented. They make exclusive use of surface data combining the information from the total signal at each triggered detector and the array geometry. We perform an analytical study of these composition estimators in order to assess their reliability, stability and possible optimization. The influence of the different slopes of the proton and Iron lateral distribution function on the discrimination power of the estimators is also studied. Additionally, the stability of the parameter in face of a possible underestimation of the size of the muon component by the shower simulation codes, as it is suggested by experimental evidence, is also studied.

\end{abstract}

 
\section{Introduction}

There are two main observation techniques of ultra-high energy cosmic rays, fluorescence and surface detection, and they both have specific composition indicators. The most reliable technique at present for composition studies is fluorescence, where the longitudinal development of the charged component of the atmospheric shower is measured. Differences in composition, manifest themselves through differences in the cross section for interactions with atmospheric nuclei. These, in turn, are mapped as different depth of maximum development of the electromagnetic component in the atmosphere ($X_{max}$) and as dispersion in the position of this maximum depth ($\Delta X_{max}$) . If proton and Iron primaries are compared, the smaller cross section of the former will produce larger  $X_{max}$ and $\Delta X_{max}$ than for protons. 

On the other hand, surface detectors make a discrete sample of the shower front at ground level, measuring the shower particles density at different distances from shower core, which is called the lateral development of the shower. Beyond a few tens of meters from the shower axis, the particle content of the shower at ground level is dominated by just two components, electromagnetic (i.e., electrons, positrons and photons) and muonic. These two sets of particles propagate in a different way through the atmosphere: the electromagnetic components propagates diffusively, while the muons do so radially from the last hadronic interaction that produce their parent mesons. Therefore, the main surface observables are related to the temporal and spatial distribution of particles. They are, for example, the slope of the lateral distribution function (LDF) used to fit the lateral development of the shower \cite{HP-Composition} \cite{VR-Composition}, the curvature of the shower front and several indicators of the time structure at a fixed point of the shower, like the rise time and fall time of the signal or their asymmetries \cite{AugerCompositonSDICRC07}. Another promising parameter is the number of muons $N_{\mu}$ \cite{Supa} (easily determined from scintillators but not from water Cherenkov tanks) because the differences between proton and Iron primaries are predicted to be around $70\%$.

In general terms, fluorescence composition indicators are regarded as easier to observe and interpret, as well as less prone to systematic errors than surface parameters do. However, fluorescence detectors suffer from a severely constrained duty cycle of approximately $10\%$ of the total time available to surface detectors. This factor alone, which makes the statistics per unit time of surface arrays an order of magnitude larger than that of fluorescence detectors, gives a great attractive to search for reliable surface composition parameters, as we do in this work.

Proton and Iron atmospheric showers are simulated using the AIRES Monte Carlo package (version 2.8.4a) \cite{Sciutto} with QGSJET-II as the hadronic interaction model. We assume a surface array of  of water Cherenkov detectors located on a triangular grid separated 1.5 km as in the Auger surface detector. The signal at each station is therefore the sum of the the electromagnetic and muonic components of the shower. The signal in each tank is measured in VEM (Vertical Equivalent Muon), i.e. the signal deposited by one vertical muon in a water Cherenkov tank, which is equivalent to 240 MeV of deposited energy \cite{Aglietta}. Therefore, to simulate the detector response easily we transform the energy of the shower particles at ground level into the measured signal considering that each muon contributes with 1 VEM and each electromagnetic particle of energy $E$, which are completely absorbed, with ($E$/240 MeV) VEM.

\section{Definition and optimization}

In the present work, we propose a new family of surface parameters defined as:
\begin{equation}
\label{Sb}
S_b = \sum_{i=1}^{N} \left[ S_{i} \times \left(\frac{r_i}{r_0}\right)^b \right] \; \; \; \; \; [\textrm{VEM}],
\end{equation}
where the sum extends over all the triggered stations \emph{N}, $r_0=1000$ m is a reference distance, $S_{i}$ is the signal in VEM measured at the $i-$th station and $r_{i}$ is the distance of this station to the shower axis in meters.

\begin{figure}[!t]
\begin{center}
\includegraphics[width=7.0cm]{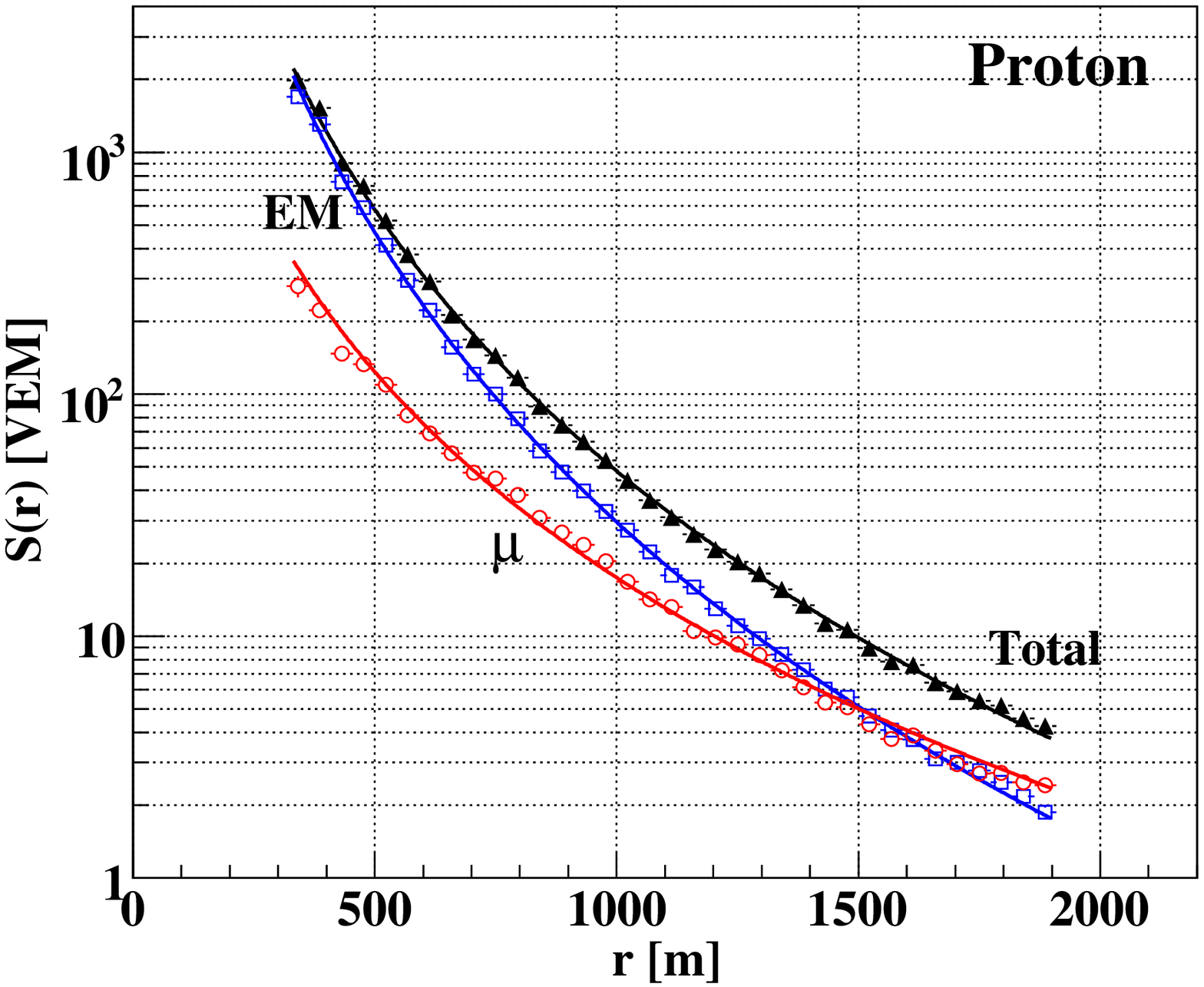}
\includegraphics[width=7.0cm]{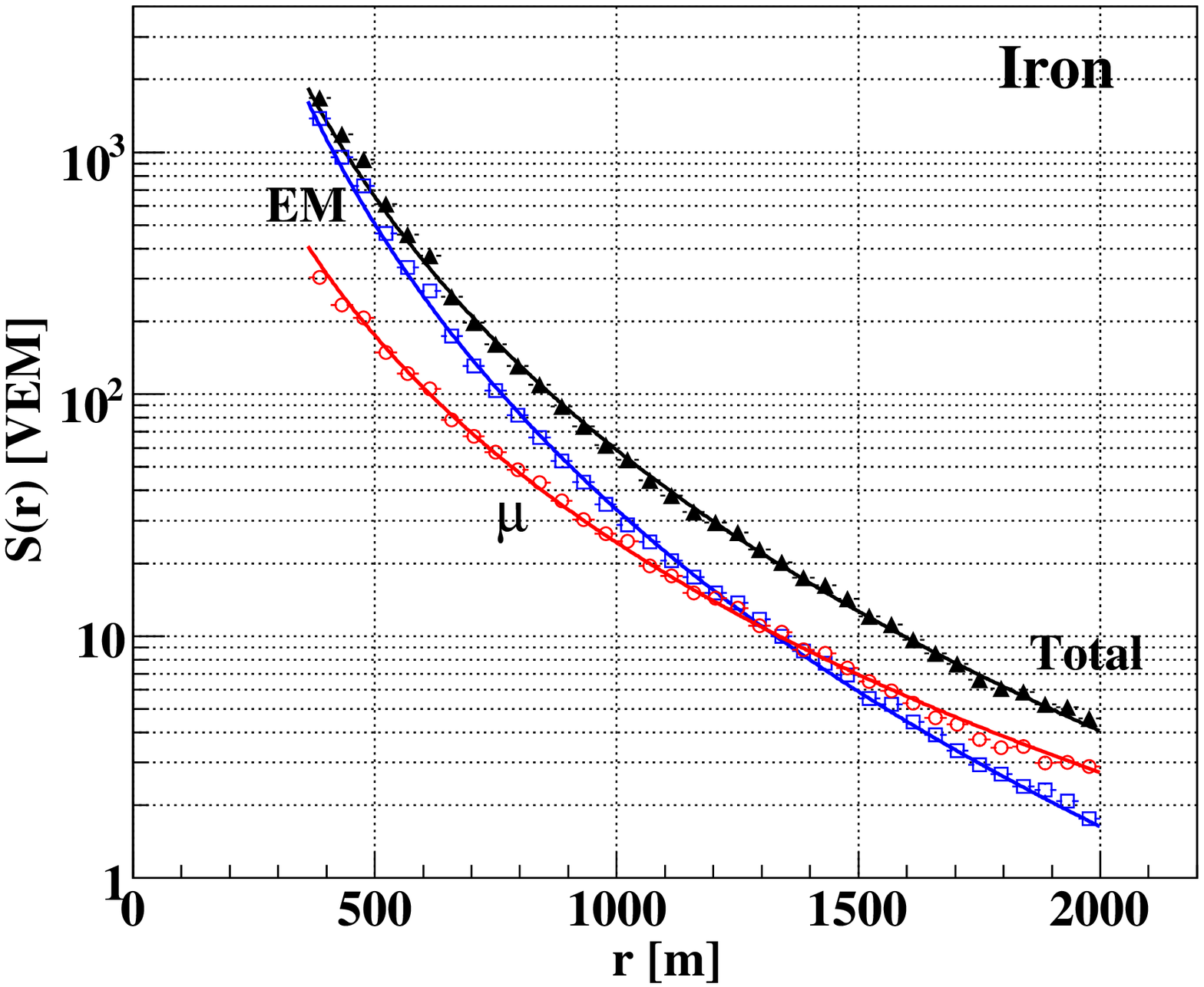}
\caption{Lateral distribution functions of the muon, electromagnetic and total signal in the Cherenkov detectors for 
simulated protons and Iron nuclei of $1 \leq \sec\theta \leq 1.2$ and $19 \leq \log(E/\textnormal{eV}) \leq 19.1$. The hadronic interaction model used to generate the showers is QGSJET-II. The solid lines correspond to the fits with a NKG-like function (see Eq. (\ref{NKGldf})). \label{FitLdfs}}
\end{center}
\end{figure}

First, we demonstrate that, under the present assumptions, the primary identity discrimination power goes through a maximum around $b \cong 3$. Different detectors or a different layout could lead, in principle, to a different optimum value for $b$. The parameter $S_b$ for a given event is constructed from the total signal in each triggered Cherenkov detector. Therefore, it depends on the normalization and shape of the lateral distribution function of the total signal. Close to the impact point of the shower, the signal is dominated by the electromagnetic particles (photons, electrons and positrons) whereas at larger distances by muons. Fig. \ref{FitLdfs} shows the muon, electromagnetic and total signal in the Cherenkov detectors as a function of the distance to the shower axis for protons and iron nuclei. The zenith angle of the simulated events considered is such that $1 \leq \sec\theta \leq 1.2$ and the primary energy $19 \leq \log(E/\textnormal{eV}) \leq 19.1$. Fig. \ref{FitLdfs} also shows the fits of the LDF of each component with a NKG-like function \cite{NKG} \cite{ICRCldf:05},
\begin{equation} 
\label{NKGldf}
S(r)=S_0 \left( \frac{r}{r_0} \right)^\beta \left( \frac{r+r_s}{r_0+r_s} \right)^\beta, 
\end{equation}
where we fix $r_s=700$ m and $r_0=1000$ m, and $S_0$ and $\beta$ are free fit parameters.
If we consider proton and Iron primaries, the discrimination power of a mass sensitive parameter $q$, like $S_b$, can be estimated by using the so-called merit factor defined as,
\begin{equation}  
\label{Eta}
\eta = \frac{E[q_{fe}]-E[q_{pr}]}{\sqrt{Var[q_{fe}]+Var[q_{pr}]}},
\end{equation}
where $E[q_A]$ and $Var[q_A]$ are the mean value and the variance, respectively, of the distribution function of 
parameter $q_A$ with $A=pr, fe$.
Assuming that the fluctuations of the total signal in an Auger Cherenkov detector are Gaussian, the distribution 
function for a given configuration of triggered stations is given by
\begin{eqnarray}
\label{Dist}
&& P(s_1, \ldots ,s_N; r_1, \ldots ,r_N) =\\ 
&& \frac{f(r_1,\ldots,r_N)}{(2\pi)^{N/2} \prod^{N}_{i=1}\sigma[S(r_i)]} \exp\left[-\sum^{N}_{i=1} \frac{(s_i-S(r_i))^2}{2\ \sigma^2[S(r_i)]} \right] \nonumber
\end{eqnarray}
where $r_i$ is the distance to the shower axis of the $i$th station (the first station, $r_1$, is the closest one), 
$S(r_i)$ is the average LDF evaluated at $r_i$, $\sigma[S(r_i)]=1.06\ [S(r_i)/\textnormal{VEM}]^{1/2}$ VEM 
\cite{ICRCldf:05} and $f(r_1,\ldots,r_N)$ is the distribution function of the distance of the different stations to the shower axis. Note that just two of the random variables $\{r_1, \ldots, r_N\}$ are independent, for instance, choosing to $r_1$ and $r_2$ (the first and second closest stations) as the independent ones, we can write 
$f(r_1,\ldots,r_N)=f_{1,2}(r_1,r_2)\delta(r_3-r_3(r_1,r_2))\ldots\delta(r_N-r_N(r_1,r_2))$, where $\delta(x)$ is the 
Dirac delta function. 

From Eqs. (\ref{Sb}) and (\ref{Dist}) we obtain the expressions for the mean value and the variance of $S_b$,
\begin{eqnarray*}
\label{MVSb}
E[S_b] &=& \sum_{i=1}^N E\left[ S(r_i) \left( \frac{r_i}{r_0}  \right)^b \right], \\
Var[S_b] &=& 1.06^2\ \sum_{i=1}^N E\left[ S(r_i) \left( \frac{r_i}{r_0}  \right)^{2 b} \right]+  \\%
&&\sum_{i=1}^N \sum_{j=1}^N cov\left[ S(r_i) \left( \frac{r_i}{r_0}  \right)^b, S(r_j) \left( \frac{r_j}{r_0} \right)^b \right], 
\end{eqnarray*}
where,
\begin{eqnarray*}
\label{Int}
E\left[ S(r_i) \left( \frac{r_i}{r_0}  \right)^x \right] = \int dr_i\ S(r_i) \left( \frac{r_i}{r_0} \right)^x f_i(r_i),\;\;\;\;\;\;\;\; \\
cov\left[ S(r_i) \left( \frac{r_i}{r_0}  \right)^b\!, S(r_j) \left( \frac{r_j}{r_0} \right)^b \right] =\;\;\;\;\;\;\;\;\;\;\;\;\;\;\;\;\;\;\;\;\;\;\;\;\;\;\; \\
\;\;\;\;\;\;\;\; \int dr_i\ dr_j\ S(r_i) \left( \frac{r_i}{r_0} \right)^b S(r_j) \left( \frac{r_j}{r_0} \right)^b f_{i,j}(r_i,r_j). 
\end{eqnarray*}
Here $f_i(r_i)$ is the distribution function of the distance to the shower axis for the $i$th station and $f_{i,j}(r_i,r_j)$ is the distribution function of the distance to the shower axis of the $i$th and $j$th stations,
\begin{eqnarray*}
\label{Fr1r2}
f_{i,j}(r_i,r_j) &=& \int dr_1 \ldots dr_{i-1} dr_{i+1} \ldots dr_{j-1}   \\
&& dr_{j+1} \ldots dr_N\ f(r_1,\ldots,r_N). 
\end{eqnarray*}
In order to simplify the expressions for the mean and variance of $S_b$ we perform the following approximations,
\begin{eqnarray*}
\label{App}
E[g(r_i)] &\cong& g(E[r_i]), \\
cov[g(r_i), g(r_j)] &\cong& \left. \frac{dg}{dr} \right|_{E[r_i]} \left. \frac{dg}{dr}\right|_{E[r_j]} cov[r_i,r_j] 
\end{eqnarray*}
where $g(r)=S(r) (r/r_0)^b$. 

\begin{figure}[!t]
\begin{center}
\includegraphics[width=7.5cm]{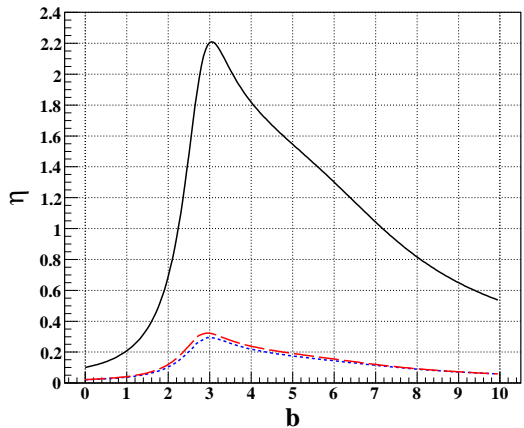}
\caption{$\eta$ as a function of $b$ for vertical showers ($1 \leq \sec\theta \leq 1.2$) and $19 \leq \log(E/eV) \leq 19.1$.
Dashed and dotted lines correspond to the cases in which the slopes of the proton and iron LDFs are the same and equal
to the corresponding to protons and iron nuclei, respectively.
\label{EtaVsB}}
\end{center}
\end{figure}

We already have analytical expressions for the average LDFs of proton and Iron primaries obtained by fitting the 
simulated data, the other ingredients needed to calculate the mean value and the variance of $S_b$ are the mean
values of the distance to the shower axis for the different stations and the covariance between all pairs of those 
random variables. We obtain these quantities from a simple Monte Carlo simulation: we uniformly distribute impact points in a triangular grid of 1.5 km of spacing, like the Auger array, and then, for each event, of zenith angle such that $\sec\theta=1.1$ and azimuthal angle uniformly distributed in $[0,2 \pi]$, we calculate the distance of each station to the shower axis. Fig. \ref{EtaVsB} shows the discrimination power $\eta$ as a function of $b$ obtained under the mentioned assumptions and simplifications. We see that $\eta$ reaches the maximum at $b\cong3$. 

\section{Modifying the slope of the LDF}

We also study the discrimination power of $S_b$ when the slope parameter $\beta$ is modified but keeping constant the 
integrated signal for distances larger than the Moliere radius, $r_M=80$ m. The modified LDF that fulfills this conditions can be written as, 
\begin{eqnarray*}
\label{LDFAC}
S(r, \beta)&=&\frac{r_s^{1+2\beta}}{(r_s+r_M)^\beta}\ \times \\
&& \textrm{Beta}(-r_s/r_M,-2 (1+\beta), 1+\beta)\ S_{\beta_0}(r) 
\end{eqnarray*}
where $\textrm{Beta}(z,a,b)=\int_0^z dt\ t^a (1-t)^b$ and $S_{\beta_0}(r)$ is the LDF of Eq. (\ref{NKGldf}) with 
the parameters $S_0$ and $\beta_0$ originally obtained from the fits. As a first calculation, fig. \ref{EtaVsB} 
also shows $\eta$ as a function of $b$ for the cases in which the slope parameter $\beta$ of both LDFs is the same 
and equal to the corresponding to protons (dashed line) and Iron (dotted line). We see that $\eta$ is considerable 
reduced when the slopes are equal.

\begin{figure}[!t]
\begin{center}
\includegraphics[width=7.5cm]{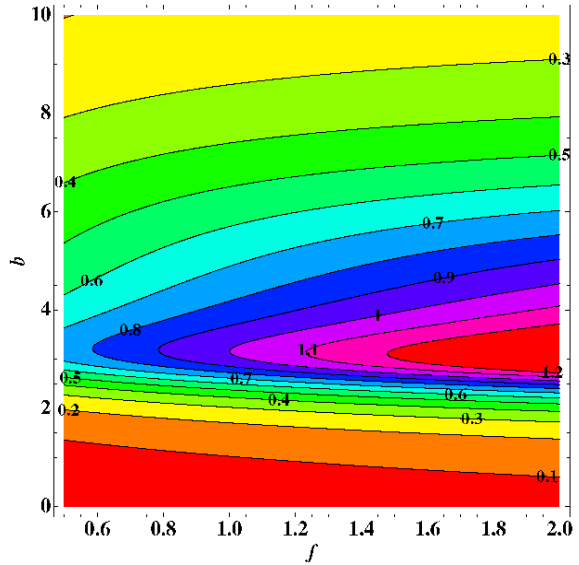}
\caption{Contour plot of $\eta(\xi,b)/\eta(1,3)$.  
\label{EtaVsDbetaB}}
\end{center}
\end{figure}

The slope of the proton LDF is smaller than the corresponding to Iron (the absolute value is grater). 
Then, we modify the slope of both LDFs such that, $\beta_{pr}(\xi)=\beta_{pr}^0-(\xi-1) \Delta \beta_0/2$ and 
$\beta_{fe}(\xi)=\beta_{fe}^0+(\xi-1) \Delta \beta_0/2$, where $\beta_{pr}^0$ and $\beta_{fe}^0$ are the proton and 
Iron slopes, respectively, obtained from the fits of the simulations, $\Delta\beta_0 = \beta_{fe}^0- \beta_{pr}^0$ 
and $\xi$ is such that $\Delta \beta(\xi)=\xi \Delta \beta_0$, i.e., $\xi=1$ corresponds to the non modified case. 
Note that for $\xi=0$, $\beta_{pr}=\beta_{fe}=(\beta_{pr}+\beta_{fe})/2$. Fig. \ref{EtaVsDbetaB} shows a contour 
plot of $\eta(\xi,b)/\eta(1,3)$ from where we see that as $\xi$ increases $\eta$ also increases. We also see that the 
maximum of $\eta$ remains close to $b=3$ almost independent of $\xi$. 

\section{Modifying the muon content of the simulated showers}

\begin{figure}[!t]
\begin{center}
\includegraphics[width=7.5cm]{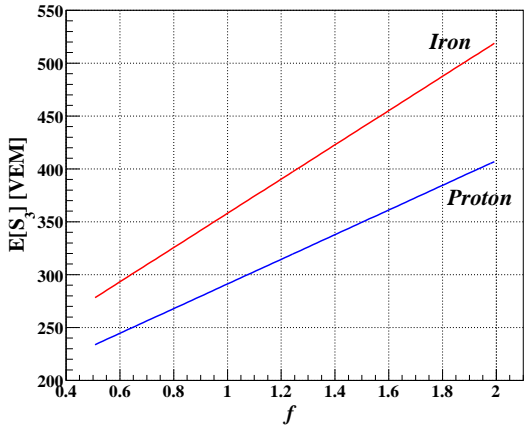}
\caption{Mean values of $S_3$ for protons and iron nuclei as a function of $f$ where $f=1$ corresponds to
the muon content predicted by QGSJET-II.
\label{S3VsFmu}}
\end{center}
\end{figure}

There is experimental evidence of a deficit in the muon content of the simulated showers \cite{AugerEngel-HM}. It is 
believed that such deficit is originated in the high energy hadronic interaction models which are extrapolations, over several orders of magnitude, of lower energy accelerator data. As mentioned, the total signal can be decomposed in the muon and electromagnetic signal. Therefore, to study how $S_b$ changes as a function of the muon content of the showers we modify the total LDFs in the following way: $S(r)=S_{em}(r)+f S_\mu(r)$ where $f$ parametrizes the 
artificial variation in the muon component. Fig. \ref{S3VsFmu} shows the mean values of $S_3$ for protons and iron 
nuclei as a function of $f$. As expected, they increases with $f$. We also see that the iron curve increases faster 
than the proton one, which means that, for larger values of $f$, the discrimination power of $S_3$ also increases. This happens because the muon content of the showers is very sensitive to the primary mass. Then, for large values of $f$ the muon component becomes more important increasing the mass sensitivity of $S_3$.

\begin{figure}[!bt]
\begin{center}
\includegraphics[width=7.5cm]{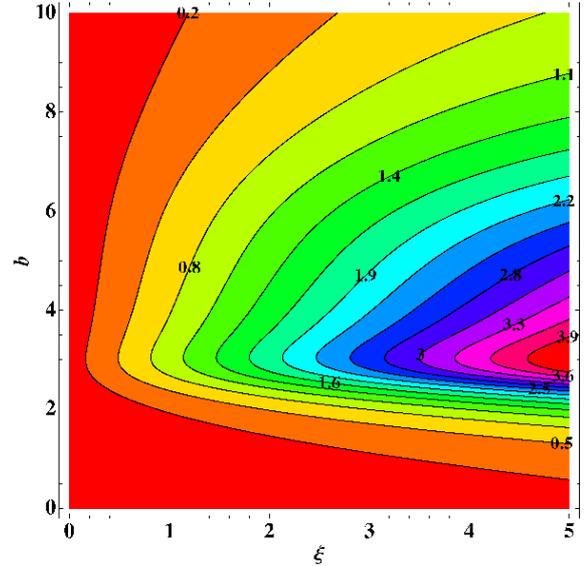}
\caption{Contour plot of $\eta(f,b)/\eta(1,3)$. $f=1$ corresponds to the muon content of the showers predicted by 
QGSJET-II.
\label{EtaVsFmuB}}
\end{center}
\end{figure}

Fig. \ref{EtaVsFmuB} shows a contour plot of $\eta(f,b)/\eta(1,3)$ from which we see how the discrimination power 
of $S_b$ increases with the muon content of the showers and that the maximum is reached at 
$b\cong3$ almost independently of $f$.

\section{Conclusions}

We propose a new family of parameters, which we call $S_b$, for composition analysis in cosmic ray surface detectors. The parameter is evaluated from the total signal and position of each triggered detector, on shower-to-shower basis. We perform an extensive analytical study of the most relevant properties of $S_{b}$. In particular, $S_b$ has been optimized to distinguish between Iron and proton primaries assuming an Auger-like water Cherenkov detectors, showing that the discrimination power between both samples is maximum for $b \cong 3$. The potential discrimination power of the $S_{3}$ parameter can be quantified by a merit factor $\eta \sim 2.2$ which is quite high if compared with the values attainable for competing parameters in current use. We have also demonstrated that, in case that the muon size is underestimated by simulation codes, as seems to be experimentally suggested,
the parameter is not only stable but improves its discrimination power. 

\section*{Acknowledgments} 

G. R. thanks to Comunidad de Madrid for a F. P. I. fellowship and to Universidad de Alcal\'a for the grant to attend to this Conference. 
G. M. T. and G. R. acknowledges the support of the ALFA-EC funds in the framework of the HELEN project. A. D. S. acknowledges a postdoctoral grant from UNAM. This work is partially supported by Spanish Ministerio de Educaci\'on y Ciencia under several projects and by Mexican PAPIIT-UNAM and CONACyT. Extensive numerical simulations were possible by the use of the UNAM super-cluster \emph{Kanbalam}.



\begin{thebibliography}{99}
   
\bibitem{HP-Composition}
 M. Ave et al. Astrop. Phys., \textbf{19}, 61 (2003).

\bibitem{VR-Composition}
 M.T. Dova et al. arXiv:astro-ph/0312463 (2003).

\bibitem{AugerCompositonSDICRC07}
M.D.Healy, for the Pierre Auger Collaboration. Proc. 30th ICRC. M\'erida, M\'exico. (2007). arXiv:0706.1569.

\bibitem{Supa}
A. D. Supanitsky et al. Astropart. Phys. \textbf{29}, 461-470 (2008).

\bibitem{Sciutto}
 S. Sciutto, AIRES user's manual and reference guide (2002). 
 http://www.fisica.unlp.edu.ar/auger/aires.

\bibitem{Aglietta}
 M. Aglietta et al. for the Pierre Auger Collaboration. Proc. of the 29t ICRC. Pune, India (2005).

\bibitem{NKG}
 K. Greisen, Progress in Cosmic Ray Physics, vol. 3, 1956.

\bibitem{ICRCldf:05}
 D. Barnhill et al., Proc. 29th ICRC \textbf{7}, 291 (2005). arXiv:astro-ph/0507590.

\bibitem{AugerEngel-HM}
 Ralph Engel, for the Pierre Auger Collaboration. Proc. 30th ICRC. M\'erida, M\'exico. (2007). arXiv:0706.1921.

 \end{thebibliography}
\end{document}